\documentclass[showkeys,aps,amsmath,amssymb,twocolumn,groupedaddress,longbibliography,floats,final,postprint,superscriptaddress]{revtex4-1}
\usepackage{enumerate}
\usepackage{graphicx}
\usepackage{amsmath,amssymb,amsthm,enumitem}

\usepackage{color}
\definecolor{LinkColor}{RGB}{199,21,133}
\usepackage[colorlinks=true,linkcolor=blue,anchorcolor=blue,citecolor=blue,urlcolor=blue]{hyperref}
\usepackage[polutonikogreek,english]{babel}
\usepackage{array}
\newcolumntype{P}[1]{>{\centering\arraybackslash}p{#1}}

\usepackage{hyperref}

\usepackage{epstopdf}
\usepackage{type1cm}
\usepackage{multirow}
\usepackage{times}
\usepackage{algpseudocode}
\usepackage{algorithm}

\def\scrR{\mathcal{R}}
\def\scrh{\mathcal{H}}

\def\scrz{\mathcal{Z}}

\def\scrj{\mathcal{C}}
\def\scro{\mathcal{O}}

\def\scrR{\mathcal{R}}
\def\scrh{\mathcal{H}}

\def\scrz{\mathcal{Z}}

\def\scrn{n}
\def\scrj{\mathcal{J}}
\def\scrd{\mathcal{D}}

\def\scrg{\mathcal{G}}

\begin{document}
\title{Classical-quantum correspondence of special and extraordinary-log criticality: Villain's bridge}
\author{Yanan Sun}
\author{Jin Lyu}
\author{Jian-Ping Lv}
\email{jplv2014@ahnu.edu.cn}
\affiliation{Anhui Key Laboratory of Optoelectric Materials Science and Technology, Key Laboratory of Functional Molecular Solids, Ministry of Education, Anhui Normal University, Wuhu, Anhui 241000, China}
\begin{abstract}
There has been much recent progress on exotic surface critical behavior, yet the classical-quantum correspondence of special and extraordinary-log criticality remains largely unclear. Employing worm Monte Carlo simulations, we explore the surface criticality at an emergent superfluid-Mott insulator critical point in the Villain representation, which is believed to connect classical and quantum O(2) critical systems. We observe a special transition with the thermal and magnetic renormalization exponents $y_t=0.58(1)$ and $y_h=1.690(1)$ respectively, which are close to recent estimates from models with discrete spin variables. The existence of extraordinary-log universality is evidenced by the critical exponent $\hat{q}=0.58(2)$ from two-point correlation and the renormalization-group parameter $\alpha=0.28(1)$ from superfluid stiffness, which obey the scaling relation of extraordinary-log critical theory and recover the logarithmic finite-size scaling of critical superfluid stiffness in open-edge quantum Bose-Hubbard model. Our results bridge recent observations of surface critical behavior in the classical statistical mechanical models [Parisen Toldin, Phys. Rev. Lett. {\bf 126}, 135701 (2021); Hu $et$ $al.$, $ibid.$ {\bf 127}, 120603 (2021); Parisen Toldin $et$ $al.$, $ibid.$ {\bf 128}, 215701 (2022)] and the open-edge quantum Bose-Hubbard model [Sun $et$ $al.$, Phys. Rev. B {\bf106}, 224502 (2022)].
\end{abstract}
\date{\today}
\maketitle

\section{Introduction}
Surface criticality (SC) refers to the critical behavior occurring on open surfaces of a critical system. For decades, SC has been a fundamental topic for modern critical theory~\cite{binder1974surface,Ohno1984,landau1989monte,diehl1997theory,pleimling2004critical,deng2005surface,deng2006bulk,Dubail2009,Hasenbusch2011,zhang2017unconventional,ding2018engineering,Weber2018,weber2019nonordinary,jian2021continuous,metlitski2020boundary}. Direct relevance has been established from SC to state-of-the-art topics including the surface effects of symmetry-protected topological phase~\cite{grover2012quantum,parker2018topological,Liu2021}, critical Casimir effects~\cite{dantchev2022critical}, boundary conformal field~\cite{cardy2004boundary,Andrei_2020}, numerical conformal bootstrap~\cite{padayasi2022extraordinary} and logarithmic critical scaling~\cite{toldin2020boundary,Hu2021,ToldinMetlitski2021extraordinary}.

The O($N$) systems---including the self-avoiding random walk ($N=0$), Ising ($N=1$), XY ($N=2$) and Heisenberg ($N=3$) models---serve as a prototypical platform for the ubiquity of criticality. Indeed, they host nontrivial SC such as the special transition and extraordinary critical phase associated with the ordinary critical phase~\cite{diehl1997theory,pleimling2004critical,deng2005surface,Hasenbusch2011,metlitski2020boundary,toldin2020boundary,Hu2021,padayasi2022extraordinary}. The characteristics of SC depend on $N$ and the space-time dimension $D=d+z$, with $d$ the spatial dimension and $z$ the dynamic critical exponent. Present work focuses on $D=3$.

\begin{figure}
\includegraphics[height=4cm,width=8cm]{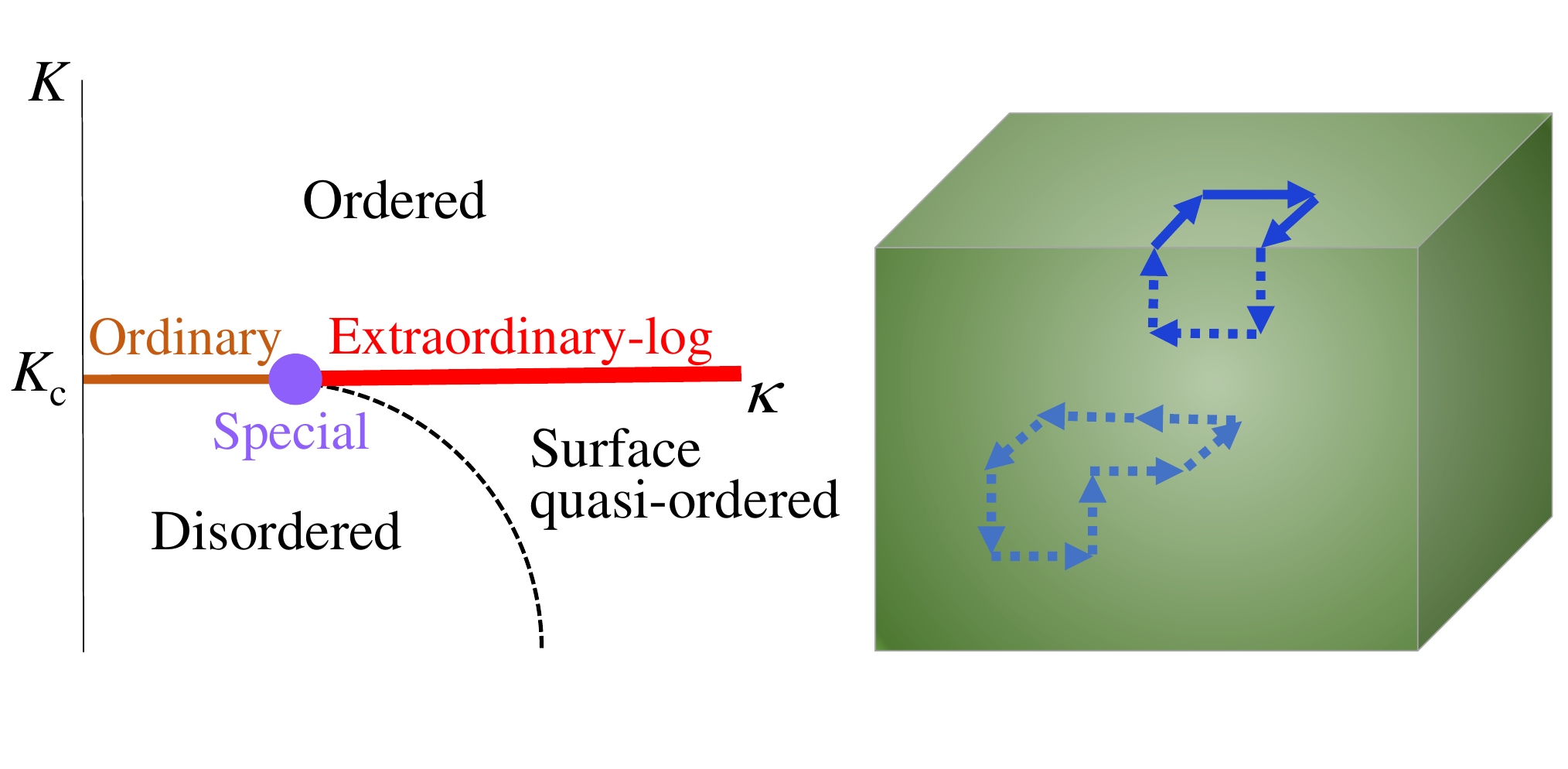}
\caption{Left panel: The phase diagram of O(2) surface criticality in terms of the bulk interaction $K$ and the ratio $\kappa$ of surface interaction enhancement~\cite{deng2005surface,metlitski2020boundary,Sun2022}. Right panel: Illustration for the open-surface Villain model, displaying two closed loops of directed flows. The directed flows on open surfaces have a distinct statistical weight from that in bulk.}~\label{guide}
\end{figure}

Figure~\ref{guide} displays the phase diagram of SC for $N=2$, where the special transition is a multi-critical point terminating the Kosterlitz-Thouless-type surface transition line and separating the ordinary and extraordinary critical phases~\cite{deng2005surface,Hu2021}. The phase diagram is therefore divided into order and disorder regimes for both surface and bulk, as well as a regime of quasi-long-range ordered surface in presence of a disordered bulk. Recently, O(2) special transitions were also found in the classical three-state Potts antiferromagnet~\cite{Zhang2022Surface} and six-state clock model~\cite{Zou2022Surface} as well as the two-dimensional quantum Bose-Hubbard model~\cite{Sun2022}---each of them can be accounted for by an {\it emergent} bulk O(2) criticality. As summarized in Table~\ref{tb1}, however, the estimates for the magnetic renormalization exponent $y_h$ from different contexts are not fully consistent.

\begin{table*}
 \begin{center}
 \caption{Universal information for the O(2) surface criticality, including the renormalization exponents $y_t$ (thermal) and $y_h$ (magnetic) for the special transition, as well as the critical exponent $\hat{q}$ and the renormalization-group parameter $\alpha$ for the extraordinary-log critical phase.}
 \label{tb1}
 \begin{tabular}{p{2cm}p{1.4cm}p{6.2cm}p{3.9cm}p{3.3cm}}
   \hline
  \hline
      \multicolumn{5}{c}{\textbf{Special transition}} \\
   \hline
   Reference & Year  &Model & $y_t$ & $y_h$ \\
   \hline

   \cite{deng2005surface}  & 2005  &  XY model & 0.608(4) & 1.675(1) \\
  \cite{Zhang2022Surface}  & 2022  & three-state antiferromagnetic Potts model &  0.59(1)    & 1.693(2) \\
  \cite{Zou2022Surface}  & 2022 & six-state clock model &   0.61(2)  & 1.688(1) \\
  present work  & 2022  & Villain model &  0.58(1) & 1.690(1)  \\
  \hline
  \hline
      \multicolumn{5}{c}{\textbf{Extraordinary-log critical phase}} \\
   \hline
  Reference  & Year    & Model &  $\hat{q}$ &  $\alpha$ \\
   \hline
   \cite{Hu2021}  & 2021   & XY model &  0.59(2)  & 0.27(2) \\
   \cite{ToldinMetlitski2021extraordinary}  & 2021     &  improved O(2) $\phi^4$ model &  & 0.300(5)\\
   \cite{Zhang2022Surface} & 2022 & three-state antiferromagnetic Potts model     & 0.60(2) &     \\
   \cite{Zou2022Surface}  &  2022 &   six-state clock model  & 0.59(1), 0.60(3), 0.59(3) &  0.26(2), 0.24(4), 0.30(3)  \\
   present work  &  2022   & Villain model  &  0.58(2) &  0.28(1) \\
  \hline
  \hline
 \end{tabular}
 \end{center}
 \end{table*}

The critical behavior of the extraordinary phase at $N=2$ has been a long-standing controversy~\cite{deng2005surface,metlitski2020boundary}.
The theory of extraordinary-log universality (ELU) was recently proposed for $2 \leq N < N_c$~\cite{metlitski2020boundary}, with $N_c$ an unknown upper bound. In this scenario, the surface two-point correlation $g(r)$ decays logarithmically with the spatial distance $r$ as~\cite{metlitski2020boundary}
\begin{equation}\label{tp1}
g(r) \sim ({\rm ln}r)^{-\hat{\eta}},
\end{equation}
where the exponent $\hat{\eta}$ merely depends on $N$. Numerical evidence for the existence of ELU has been obtained from critical Heisenberg~\cite{toldin2020boundary} and XY~\cite{Hu2021,ToldinMetlitski2021extraordinary} models. Motivated by the Fourier-mode-dependent finite-size scaling (FSS) of magnetic fluctuations~\cite{Wittmann2014,Flores-Sola2016} and the two-length scenarios in different contexts of bulk criticality~\cite{papathanakos2006finite,shao2016quantum,grimm2017geometric,zhou2018random,FangComplete,Lv2019}, an alternative scaling form of $g(r)$ was conjectured~\cite{Hu2021} for ELU. This conjecture is based on the $L$ dependence ($L$ is linear size) of critical magnetic fluctuations at zero and smallest non-zero modes, which scale as $L^2 ({\rm ln}L)^{-\hat{q}}$ and $L^2 ({\rm ln}L)^{-\hat{\eta}}$ with the exponents $\hat{q}$ and $\hat{\eta}=\hat{q}+1$, respectively. The critical scaling behavior of $g(r)$ is described by~\cite{Hu2021}
\begin{equation}\label{tp3}
g(r) \sim \begin{cases} ({\rm ln}r)^{-\hat{\eta}}, & {\rm ln}r \le  \scro [({\rm ln}L)^{\hat{q}/\hat{\eta}}],  \\
({\rm ln}L)^{-\hat{q}},  & {\rm ln}r \ge \scro [({\rm ln}L)^{\hat{q}/\hat{\eta}}].
\end{cases}
\end{equation}
For the $N=2$ case, the first result of $\hat{q}$ is $\hat{q}=0.59(2)$~\cite{Hu2021}. The coexistence of the exponents $\hat{q}$ and $\hat{\eta}$ was confirmed in the context of the ELU in three-state Potts antiferromagnet~\cite{Zhang2022Surface}. Table~\ref{tb1} lists the results of $\hat{q}$ from different contexts~\cite{ToldinMetlitski2021extraordinary,Zhang2022Surface,Zou2022Surface}. Recall the scaling formula proposed~\cite{metlitski2020boundary} for the helicity modulus $\Upsilon$, which measures the response of a system to a twist in boundary conditions~\cite{fisher1973}. The FSS of $\Upsilon$ is written as
\begin{equation}\label{qp1}
\Upsilon L \sim 2 \alpha ({\rm ln}L)
\end{equation}
with the universal renormalization-group parameter $\alpha$. Further, the scaling relation between $\hat{q}$ and $\alpha$ reads~\cite{metlitski2020boundary}
\begin{equation}\label{qp2}
\hat{q}=\frac{N-1}{2 \pi \alpha}.
\end{equation}
This relation has been verified for critical Heisenberg~\cite{toldin2020boundary} and XY~\cite{Hu2021} models as well as an emergent O(2) critical point~\cite{Zou2022Surface} (Table~\ref{tb1}).

Despite the complementary evidence for classical ELU and the numerous efforts toward a quantum counterpart, the self-contained picture for classical-quantum correspondence remains badly awaited~\cite{metlitski2020boundary}. Motivated by the exotic surface effects of symmetry-protected topological phases, SC has been extensively studied in dimerized antiferromagnetic quantum Heisenberg and XXZ models~\cite{zhang2017unconventional,ding2018engineering,Weber2018,weber2019nonordinary,jian2021continuous,Weber2021,Zhu2021Exotic,ding2021special}, yet the existence of quantum ELU is still controversial. Very recently, quantum O(2) SC was explored in an open-edge Bose-Hubbard model of interacting soft-core bosons, where quantum special transition and quantum ELU were observed~\cite{Sun2022}.

To establish a direct classical-quantum correspondence of O(2) SC, we formulate an open-surface Villain (OSV) model and study the special transition and extraordinary-log critical phase. Such a methodology was applied to the linear-response dynamics at a quantum O(2) critical point~\cite{witczak2014dynamics}. The Villain model can be viewed as a variant of the quantum phase model, which is connected with the unit-filling Bose-Hubbard model. The Hamiltonian of the Bose-Hubbard model reads~\cite{fisher1989boson}
\begin{equation}\label{bosehubbard}
\scrh_{{\rm BH}} = -t \sum_{\bf \langle rr' \rangle} ({\hat \Phi}_{\bf r}^{\dagger}
         {\hat \Phi}_{\bf r'} +{\hat \Phi}_{\bf r} {\hat \Phi}_{\bf r'}^{\dagger})+\frac{U}{2} \sum_{\bf r} {\hat n}_{\bf r}^2
\end{equation}
where ${\hat \Phi}_{\bf r}^{\dagger}$ and ${\hat \Phi}_{\bf r}$ are the bosonic creation and annihilation operators at site ${\bf r}$ respectively, ${\hat n}_{\bf r}={\hat \Phi}_{\bf r}^{\dagger}{\hat \Phi}_{\bf r}$ is the particle number operator, $t$ represents the strength of nearest-neighbor hopping, and $U$ denotes onsite repulsion. The superfluid-Mott insulator transition of unit-filling Bose-Hubbard model belongs to emergent O(2) criticality~\cite{zhai2021ultracold}. By integrating out amplitude fluctuations, the quantum phase model is formally~\cite{Fisher1988}
\begin{equation}
\scrh_{{\rm QR}} = -t \sum_{\bf \langle rr' \rangle} \cos({\hat \phi}_{\bf r} - {\hat \phi}_{\bf r'})+\frac{U}{2} \sum_{\bf r} {\hat n}_{\bf r}^2,
\end{equation}
where ${\hat n}_{\bf r}$ is now the deviation from mean filling and $t$ is a multiple of that in Eq.~(\ref{bosehubbard}). ${\hat \phi}_{\bf r}$ is conjugate to ${\hat n}_{\bf r}$ by ${\hat n}_{\bf r}=(1/i) (\partial/\partial {\hat \phi}_{\bf r})$. Hence, the quantum phase model is rewritten in angle representation as~\cite{Wallin1994}
\begin{equation}
\scrh_{{\rm QR}} = -t \sum_{\bf \langle rr' \rangle} \cos({\phi}_{\bf r} - {\phi}_{\bf r'})+\frac{U}{2} \sum_{\bf r} (\frac{1}{i}\frac{\partial }{\partial {\phi}_{\bf r}})^2.
\end{equation}
Using standard Suzuki-Trotter decomposition, the inverse temperature $\beta$ is divided into slices with width $\Delta\tau$ and a path-integral representation can be established~\cite{Wallin1994}. Further, the Villain approximation is performed for $\cos(\phi)$ term, which is reexpressed by periodic Gaussians as
$\exp({t\Delta\tau\cos({\phi})}) \rightarrow  \exp({t\Delta\tau})\sum_n\exp({-\frac{1}{2}t\Delta\tau
({\phi} - 2\pi n)^2})$ with $n$ an integer, hence the periodicity in $\phi$ is unaffected~\cite{villain75}. Finally, by employing Poisson summation, it can be shown that the ground-state energy equals the free energy of the classical Hamiltonian~\cite{Wallin1994,kisker1997two}
\begin{equation}\label{vlm}
\scrh_{{\rm V}}=\frac{1}{2K} \sum^{\Delta\scrj=0}_{\bf \langle rr' \rangle} \scrj^2_{{\bf rr'}},
\end{equation}
where the parameter $K$ relates to the ratio $t/U$. $\scrj_{{\bf rr'}} \in \{...,-2,-1,0,1,2,...\}$ parameterizes the integer-valued directed flow between nearest-neighbor sites ${\bf r}$ and ${\bf r'}$. $\Delta\scrj=0$ denotes the absence of source and sink for flows---$\forall \, {\bf r}$, $\scrd_{\bf r}=\sum_{\bf r'} \scrj_{{\bf rr'}}=0$. The model (\ref{vlm}) harbors the superfluid-Mott insulator transition~\cite{cha1991universal,Wallin1994,alet2003cluster,vsmakov2005universal,chen2014universal,witczak2014dynamics}, while a rigorous analysis for massless bulk phase became available recently~\cite{dario2020massless}.

Recall the hopping enhancement on open edges of quantum Bose-Hubbard model~\cite{Sun2022}. Here, we formulate an OSV model, where the parameter $K$ becomes tunable on open surfaces. Hence, the OSV model is a classical counterpart of open-edge quantum Bose-Hubbard model and a possible testbed for classical-quantum correspondence. Besides, the OSV model admits state-of-the-art worm Monte Carlo simulations, by which the correlation function and superfluid (SF) stiffness can be efficiently sampled.

\section{Model}
The Hamiltonian of the OSV model reads
\begin{equation}\label{HamOV}
\scrh_{{\rm OSV}}=\sum^{\Delta\scrj=0}_{\bf \langle rr' \rangle} \frac{\scrj^2_{{\bf rr'}}}{2K_{{\bf rr'}}},
\end{equation}
where the parameter $K_{\bf rr'}$ is for the nearest-neighbor sites ${\bf r}$ and ${\bf r'}$ on simple-cubic lattices. We impose open boundary conditions along [001] ($z$) direction as well as periodic boundary conditions along [100] ($x$) and [010] ($y$) directions. Hence, there are a pair of open surfaces. We set $K_{\bf rr'}=K'$ if ${\bf r}$ and ${\bf r'}$ are on the same open surface and $K_{\bf rr'}=K_c=0.333\,067\,04$ for other situations, with $K_c$ the bulk critical point of model~(\ref{vlm}) determined previously by two of us and coworkers~\cite{Xu2019}. The surface enhancement of $K_{\bf rr'}$ is parameterized by $\kappa=(K'-K_c)/K_c$. A directed-flow state for model (\ref{HamOV}) is illustrated by Fig.~\ref{guide}.

\section{Methodology based on a worm Monte Carlo algorithm}
We simulate model~(\ref{HamOV}) with the side length $L$ of simple-cubic lattice ranging from $L=4$ to $256$. To this end, we formulate a worm Monte Carlo algorithm along the lines of Ref.~\cite{prokof2001worm}. Similar formulations of Monte Carlo algorithms have been applied to Villain model~\cite{alet2003cluster,Xu2019,Lyu2022} and other lattice models~\cite{Deng2007,zhang2009worm,lv2011worm}. Here, the methodology contains three components: extending state space (Sec.~\ref{esp}), update scheme (Sec.~\ref{rwd}) and sampling of quantities (Sec.~\ref{moq}).

Conclusions of present work are drawn on the basis of FSS analyses of Monte Carlo data, for which we employ least-squares fits. In the fits, we analyze the dependence of the residuals chi$^2$ on the cut-off size $L_{\rm min}$. In principle, the reasonable fit corresponds to
the smallest $L_{\rm min}$ for which chi$^2$ per degree of freedom (DoF) obeys ${\rm chi}^2/{\rm DoF}=\scro(1)$ and for which subsequently increasing $L_{\rm min}$ do not induce a decrement of ${\rm chi}^2/{\rm DoF}$ over a unit. Practically, by ``reasonable" one means that ${\rm chi}^2/{\rm DoF} \approx 1$.

\subsection{Extending state space}~\label{esp}
The partition function of model~(\ref{HamOV}) reads
\begin{equation}
\scrz_{{\rm OSV}}= \sum_{\Delta\scrj=0}  \prod_{\langle {\bf r}
{\bf r'}\rangle}  e^{-\frac{\scrj^2_{{\bf r}{\bf r'}}}{2K_{{\bf r}{\bf r'}}}},
\end{equation}
where the summation runs over states in the directed-flow state space. For later convenience, $\scrz_{{\rm OSV}}$ is unbiasedly reformulated in an extended state space as
\begin{equation}\label{3det}
\scrz'_{{\rm OSV}}=\frac{1}{L^3} \sum_{\Delta\scrj=0; \, \{I, M\}_{\rm 3d}} \delta_{I,M} \prod_{\langle {\bf r}
{\bf r'}\rangle}  e^{-\frac{\scrj^2_{{\bf r}{\bf r'}}}{2K_{{\bf r}{\bf r'}}}}
\end{equation}
or
\begin{equation}\label{2det}
\scrz''_{{\rm OSV}}=\frac{1}{L^2} \sum_{\Delta\scrj=0; \, \{I, M\}_{\rm 2d}} \delta_{I,M} \prod_{\langle {\bf r}
{\bf r'}\rangle}  e^{-\frac{\scrj^2_{{\bf r}{\bf r'}}}{2K_{{\bf r}{\bf r'}}}}
\end{equation}
by including two additional degrees of freedom---in a state, the sites $I$ and $M$ are specified on the whole lattice [Eq.~(\ref{3det})] or an open surface [Eq.~(\ref{2det})]. The summations run over the states in extended state spaces. $\delta$ denotes the Kronecker delta function.

The simulated partition functions in extended state space read
\begin{equation}~\label{3desp}
\scrz'_{\rm sim}=\scrz'_{{\rm OSV}}+\lambda \scrg'
\end{equation}
and
\begin{equation}~\label{2desp}
\scrz''_{\rm sim}=\scrz''_{{\rm OSV}}+\lambda \scrg''
\end{equation}
with
\begin{equation}
\scrg'=\frac{1}{L^3} \sum_{\Delta\scrj=0; \, \{I, M\}_{\rm 3d}} (1-\delta_{I,M}) \prod_{\langle {\bf r}
{\bf r'}\rangle}  e^{-\frac{\scrj^2_{{\bf r}{\bf r'}}}{2K_{{\bf r}{\bf r'}}}}
\end{equation}
and
\begin{equation}
\scrg''=\frac{1}{L^2} \sum_{\Delta\scrj=0; \, \{I, M\}_{\rm 2d}} (1-\delta_{I,M}) \prod_{\langle {\bf r}
{\bf r'}\rangle}  e^{-\frac{\scrj^2_{{\bf r}{\bf r'}}}{2K_{{\bf r}{\bf r'}}}}
\end{equation}
respectively, where $\lambda$ is tunable. The subspaces with $I \ne M$, denoted in following by $S'$ and $S''$, contribute to $\scrg'$ and $\scrg''$, respectively.

\subsection{Update scheme}~\label{rwd}
To simulate partition function (\ref{3desp}), an update scheme can be designed through a biased random walk that obeys detailed balance, by moving $I$ and $M$ on simple-cubic lattice. The procedure starts with $I=M$ in original state space.
As $I$ ($M$) moves to a neighbor $I_\scrn$ ($M_\scrn$), the flow on edge $II_\scrn$ ($MM_\scrn$) will be updated by adding a unit directed flow from $I$ to $I_\scrn$ ($M_\scrn$ to $M$). Such a movement continues. When $I \ne M$, the flows passing $I$ and $M$ are not conserved, i.e., $\scrd_I \neq 0$ and $\scrd_M \neq 0$, and $S'$ space is hit. When $I=M$, the original state space is hit again. Thus, a movement of $I$ or $M$ is either a step of random walk in $S'$ space or between $S'$ and original spaces. More precisely, a Monte Carlo microstep is described in \textbf{Algorithm} \textbf{1}.

\begin{algorithm}[H]
\caption{Global update.}
\label{alg:p1}
\begin{algorithmic}
\item 1. If $I=M$, randomly and uniformly choose a new site $I'$ and set $I=M=I'$, ${\rm sign}(I)=1$, ${\rm sign}(M)=-1$.
\item 2. Interchange $I \leftrightarrow M$ and ${\rm sign}(I) \leftrightarrow {\rm sign}(M)$ with probability $1/2$.
\item 3. If $I$ is on an open surface, exit present micro Monte Carlo step with the probability $1/6$~\footnote{Practically, this can be realized alternatively by setting pseudo edges between the two open surfaces, which have zero relative statistical weights for finite $\scrj$ with respect to $\scrj=0$}.
\item 4. Randomly and uniformly choose a neighbor $I_\scrn$ of $I$.
\item 5. Propose to move $I \rightarrow I_\scrn$ by updating the flow $\scrj_{II_\scrn}$ to $\scrj'_{II_\scrn}$:
\begin{equation}
\scrj'_{II_\scrn}=\scrj_{II_\scrn}+{\rm sign}(I \rightarrow I_\scrn){\rm sign}(I), \nonumber
\end{equation}
where ${\rm sign}(I \rightarrow I_\scrn)=\pm 1$, permanently parametrizing the flow direction along edge-$II_\scrn$.
\item 6. Accept the proposal with probability
\begin{equation}
\label{eq6}
P_{\rm acc} =\min[1, e^{\frac{-(\scrj'^2_{II_\scrn}-\scrj^2_{II_\scrn})}{2K'}}] \nonumber
\end{equation}
if $I$ and $I_\scrn$ are on an open surface, and, for other situations, with
\begin{equation}
\label{eq6}
P_{\rm acc} = \min[1, e^{\frac{-(\scrj'^2_{II_\scrn}-\scrj^2_{II_\scrn})}{2K}}]. \nonumber
\end{equation}
\end{algorithmic}
\end{algorithm}

In line with partition function (\ref{2desp}), we formulate a {\it supplementary} procedure to \textbf{Algorithm} \textbf{1} by the random walk of $I$ and $M$ on a specified open surface, which is described in \textbf{Algorithm} \textbf{2}. We emphasize that \textbf{Algorithm} \textbf{2} itself is not ergodic.

\begin{algorithm}[H]
\caption{Restricted update.}
\label{alg:p2}
\begin{algorithmic}
\item 1. If $I=M$, randomly and uniformly choose a new site $I'$ on a specified open surface and set $I=M=I'$, ${\rm sign}(I)=1$, ${\rm sign}(M)=-1$.
\item 2. Interchange $I \leftrightarrow M$ and ${\rm sign}(I) \leftrightarrow {\rm sign}(M)$ with probability $1/2$.
\item 3. Randomly and uniformly choose a neighbor $I_\scrn$ on the same open surface of $I$.
\item 4. Propose to move $I \rightarrow I_\scrn$ by updating the flow $\scrj_{II_\scrn}$ to $\scrj'_{II_\scrn}$:
\begin{equation}
\scrj'_{II_\scrn}=\scrj_{II_\scrn}+{\rm sign}(I \rightarrow I_\scrn){\rm sign}(I). \nonumber
\end{equation}
\item 5. Accept the proposal with probability
\begin{equation}
\label{eq6}
P_{\rm acc} =
\min[1, e^{\frac{-(\scrj'^2_{II_\scrn}-\scrj^2_{II_\scrn})}{2K'}}]. \nonumber
\end{equation}
\end{algorithmic}
\end{algorithm}

A closed loop of directed flow is superposed once $I$ meets $M$, and closed loops can be consecutively superposed. The update scheme switches between \textbf{Algorithm} \textbf{1} and \textbf{2}, when a fixed number of closed loops is generated.

Practically, parallel simulations are carried out and a large number of closed loops are created. Around the special transition point ($0.44 \leq \kappa \leq 0.4428$), the number of generated closed loops ranges from $2.03 \times 10^{10}$ to $5.42 \times 10^{11}$ for $8 \leq L \leq 128$ and increases to $5.06 \times 10^{11}$ at $L=256$. In the deep extraordinary critical regime ($\kappa=5$ and $10$), the number ranges from $4.95 \times 10^{9}$ to $7.93 \times 10^{10}$ for $8 \leq L \leq 128$ and reaches $5.28 \times 10^{10}$ at $L=256$. For each independent simulation, the initial one sixth of closed loops are used for thermalization.

\subsection{Sampling of quantities}~\label{moq}
\textit{Extended state space.} Using \textbf{Algorithm} \textbf{2}, we sample the probability distribution of the distance between $I$ and $M$, which is an unbiased estimator for the surface two-point correlation $g(r_1,r_2)$ [$g(0,0)\equiv 1$]. In particular, we define
\begin{equation}
G_1=[g(0,L/4)+g(L/4,0)]/2
\end{equation}
and
\begin{equation}
G_2=[g(0,L/2)+g(L/2,0)]/2.
\end{equation}
The surface susceptibility $\chi$ can be evaluated by the number $n_s$ of worm steps between subsequent hits to the original state space. Accordingly, $\chi$ is defined by
\begin{equation}\label{defineTw}
\chi= \langle n_s \rangle.
\end{equation}

\textit{Original state space.} The following quantities are sampled in original state space. First, the winding probabilities are given by
\begin{eqnarray}
R_x &=&  \langle \scrR_x \rangle = \langle \scrR_y \rangle, \\
R_a &=& \langle 1 -  (1-\scrR_x)(1-\scrR_y) \rangle,  \\
R_2 &=& \langle \scrR_x \scrR_y \rangle,
\end{eqnarray}
for which $\scrR_\alpha=1$ (resp. $\scrR_\alpha=0$) corresponds to the event that directed flows wind (resp. do not wind) in the $\alpha$ direction of simple-cubic lattice. Hence, $R_x$, $R_a$ and $R_2$ define the probabilities that the winding of directed flows exists in $x$ direction, in at least one direction and in both $x$ and $y$ directions, respectively. More or less similar dimensionless quantities can also be defined for geometric percolation transitions~\cite{langlands1992universality,pinson1994critical,ziff1999shape,Wang2013,Xu2014,Hu2020,Wang2021}. The SF stiffness relates to winding number fluctuations as
\begin{equation}
\rho=\langle \mathcal{W}_x^2+\mathcal{W}_y^2 \rangle/(2L),
\end{equation}
with $\mathcal{W}_x$ and $\mathcal{W}_y$ the winding numbers in $x$ and $y$ directions, respectively.

Further, for an observable (say $\scro$), we define its covariance with the surface energy $\varepsilon_s$ as
 \begin{equation}
  C_{O}= \frac{1}{K'^2} (\langle \scro \varepsilon_s \rangle - \langle \scro \rangle \langle \varepsilon_s \rangle)
 \label{eq:Rp}
 \end{equation}
with
 \begin{equation}
	\varepsilon_s = \frac{1}{2} \sum \limits_{\langle {\bf rr'} \rangle_{s}} \scrj^2_{{\bf rr'}},
	\label{ener}
 \end{equation}
where the summation runs over edges on an open surface. Accordingly, $C_{O}$ equals to the derivative of $O=\langle \scro \rangle$ with respect to $K'$.

\section{Special transition}
\subsection{Location}\label{dtc}
\begin{figure}
\includegraphics[height=10cm,width=8cm]{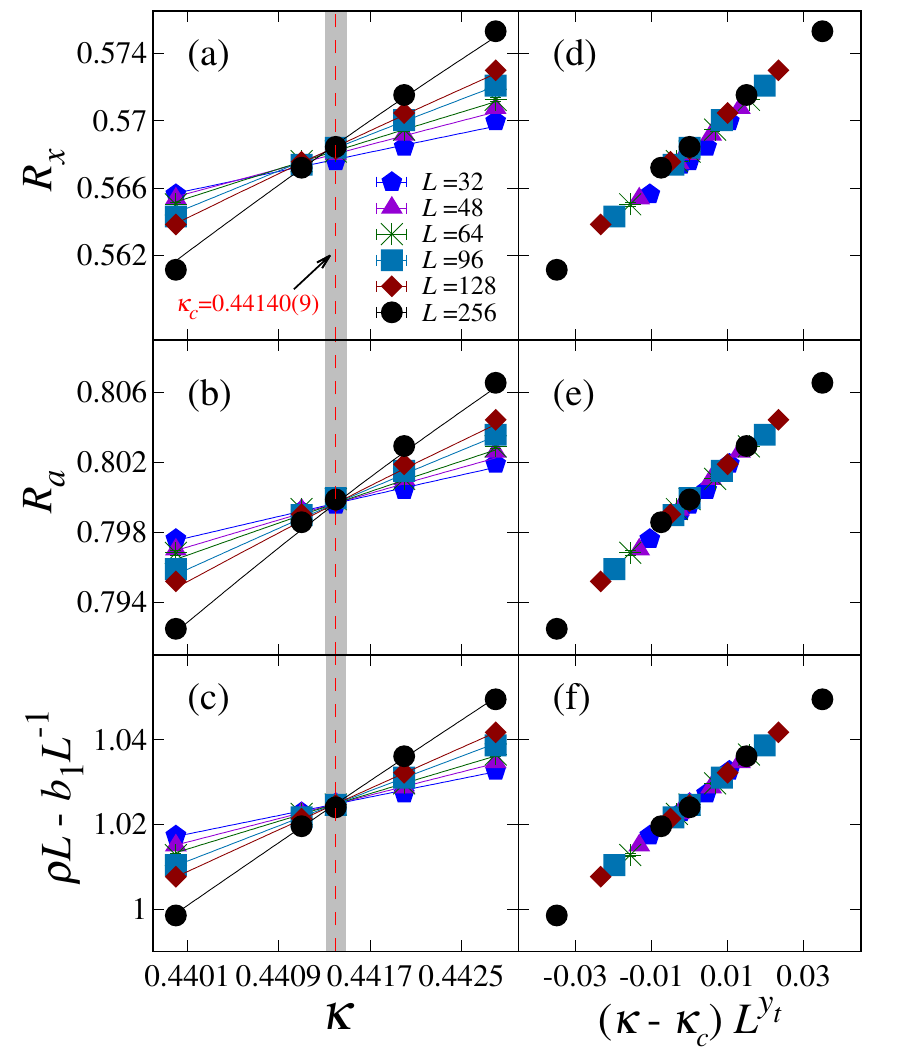}
\caption{The winding probabilities $R_x$ (a) and $R_a$ (b) and the scaled SF stiffness $\rho L-b_1 L^{-1}$ (c) versus $\kappa$, where $b_1=-0.329$ is taken from a preferred least-squares fit to Eq.~(\ref{fitrhos1}). In panels (a), (b) and (c), the symbols stand for Monte Carlo data and the lines are drawn according to preferred fits. In panels (d), (e) and (f), the horizontal coordinates are set to be $(\kappa-\kappa_c)L^{y_t}$, with $\kappa_c=0.44140$ and $y_t=0.58$.}\label{spe1}
\end{figure}

We locate the special transition by varying $\kappa$. Recall the application of dimensionless winding probabilities in flow representation for O(2) criticality~\cite{Xu2019,Lyu2022} as well as an analog in world-line representation for the quantum special transition of Bose-Hubbard model~\cite{Sun2022}. When a special transition occurs at $\kappa_c$, $R$ ($R=R_x, R_a$) is assumed to scale as
\begin{equation}\label{RrGc1}
R=\tilde{R}(\epsilon L^{y_t})
\end{equation}
around $\kappa_c$, where $\epsilon$ equals $\kappa-\kappa_c$, $y_t$ denotes the thermal renormalization exponent, and $\tilde{R}$ is a scaling function. Figures~\ref{spe1}(a) and (b) respectively show $R_x$ and $R_a$ versus $\kappa$ for $L=32$, $48$, $64$, $96$, $128$ and $256$. Scale invariance is observed at $\kappa_c \approx 0.441$. A more precise result comes from least-squares fits of the Monte Carlo data to the expansion of Eq.~(\ref{RrGc1})
\begin{equation}\label{RrGc2}
 R=R^{*}+ a_1 \epsilon L^{y_t}+ a_2 \epsilon^2 L^{2y_t}+b_1 L^{-\omega_1}+...
\end{equation}
where $R^{*}$ is a critical value, $a_1$, $a_2$ and $b_1$ are non-universal constants, and $b_1 L^{-\omega_1}$ represents the leading finite-size corrections with correction exponent $\omega_1$. For $R_x$, when the four terms in right-hand side of Eq.~(\ref{RrGc2}) are all included, preferred fits with ${\rm chi}^2/{\rm DoF} \approx 0.7$ are achieved and yield $0.44141(5)$ and $0.44140(8)$ for $L_{\rm min}=8$ and $16$, respectively. Meanwhile, we obtain the estimates of $\omega_1$ as $\omega_1 = 1.06(9)$ and $1.1(3)$. A close value of leading correction exponent---$\omega_1=1$ from irrelevant surface fields---has been applied to the special transitions with $N=1$~\cite{Hasenbusch2011} and $N=3$~\cite{toldin2020boundary}. Despite these observations, for caution, we should be aware of the correction exponent $\omega_1 \approx 0.789$ originating from O(2) bulk irrelevant field~\cite{guida1998}. A useful procedure is to increase $L_{\rm min}$ gradually and monitor the stability of fitting results. In this process, the finite-size corrections become more and more negligible. When $\omega_1=1$ is fixed, we obtain $\kappa_c=0.44144(3)$, $0.44143(4)$, $0.44141(5)$, $0.44142(7)$ and $0.44143(9)$ with $L_{\rm min}=8$, $16$, $32$, $48$ and $64$, respectively. In Fig.~\ref{spe1}(b), the finite-size corrections for $R_a$ are relatively weak; hence, we perform fits without incorporating any finite-size correction. Stable results are achieved for large $L_{\rm min}$. In particular, we obtain $\kappa_c=0.44132(3)$, $0.44137(3)$, $0.44139(4)$, $0.44138(5)$ and $0.44140(6)$ for $L_{\rm min}=32$, $48$, $64$, $96$ and $128$ respectively, with $0.5 \lessapprox {\rm chi}^2/{\rm DoF} \lessapprox 1.0$.

\begin{table}
\begin{center}
\caption{Fits of the winding probabilities $R_x$ and $R_a$ to Eq.~(\ref{RrGc2}) and the SF stiffness $\rho$ to Eq.~(\ref{fitrhos1}) for the special transition. ``$\mathcal{Q}$'' is the abbreviation of sampled quantity and ``---'' indicates that the corresponding term is not included in fitting.}
\label{TABLE_S}
\begin{tabular}{p{0.6cm}p{0.7cm}p{1.3cm}p{1.6cm}p{1.1cm}p{1.6cm}p{0.9cm}}
\hline
\hline
$\mathcal{Q}$ &$L_{\rm min}$  & ${\rm chi}^2$/DoF & $\kappa_c$&$y_t$&$R^*$ or $a_0$ &$\omega_1$\\
\hline
\multirow{7}{*}{$R_x$}
&8  &23.81/33&0.44141(5)&0.59(1) &0.5687(2)&1.06(9)\\
&16  &18.49/28&0.44140(8)&0.58(1) &0.5686(4)&1.1(3)\\
&8  &24.23/34&0.44144(3)&0.59(1) &0.56884(8)&1\\
&16  &18.61/29&0.44143(4)&0.58(1) &0.5688(1)&1\\
&32  &17.35/24&0.44141(5)&0.58(2) &0.5687(2)&1\\
&48  &11.19/19&0.44142(7)&0.59(2) &0.5687(3)&1\\
&64 &9.87/14&0.44143(9)&0.60(2) &0.5688(4)&1\\
\hline
\multirow{5}{*}{$R_a$}
&32  &25.48/25&0.44132(3)&0.58(2) &0.79966(7)&---\\
&48  &14.99/20&0.44137(3)&0.58(2) &0.79980(9)&---\\
&64  &7.30/15&0.44139(4)&0.61(2) &0.7999(1)&---\\
&96  &5.27/10&0.44138(5)&0.62(3) &0.7998(2)&---\\
&128  &3.98/5&0.44140(6)&0.61(4) &0.7999(2)&---\\
\hline
\multirow{5}{*}{$\rho$}
&8  &45.76/34&0.44132(2)&0.59(1) &1.0235(3)&1\\
&16  &20.35/29&0.44141(3)&0.58(1) &1.0248(4)&1\\
&32  &16.40/24&0.44144(5)&0.58(1) &1.0253(7)&1\\
&48  &9.41/19&0.44146(6)&0.57(2) &1.026(1)&1\\
&64  &8.69/14&0.44146(8)&0.57(2) &1.026(2)&1\\
\hline
\hline
\end{tabular}
\end{center}
\end{table}

\begin{table}
\begin{center}
\caption{Fits of the covariances $C_{R_x}$, $C_{R_a}$, $C_{R_2}$ and $C_{\rho L}$ to Eq.~(\ref{fitco}) at the special transition.}
\label{TABLE_Co}
\begin{tabular}{p{0.8cm}p{0.8cm}p{1.3cm}p{1.4cm}p{1.5cm}p{1.3cm}p{0.35cm}}
\hline	
\hline		
$\mathcal{Q}$ &$L_{\rm min}$ &  ${\rm chi}^2$/DoF &$a_0$&$y_t$ &$b_1$&$\omega_1$ \\
\hline
\multirow{6}{*}{$C_{R_x}$}
& 4 & 11.22/6 & 0.623(3) & 0.570(1)& -0.13(1) & 1 \\
&8    &6.68/5&0.611(6)&0.574(2)&-0.08(3)&1\\
&16     &0.79/4&0.64(1)&0.566(4)&-0.28(9)&1\\
&16    &10.84/5&0.600(3)&0.578(1) &---&---\\
&32    &2.06/4&0.612(5)&0.573(2)&---&---\\
&48   &0.42/3&0.619(8)&0.570(3)&---&---\\
\hline
\multirow{6}{*}{$C_{R_a}$}
&4 & 24.58/6 & 0.642(4) & 0.563(1) & -0.25(1) & 1 \\
&8    &4.57/5&0.615(7)&0.572(3)&-0.11(3)&1\\
&16    &1.07/4&0.64(1)&0.565(4)&-0.3(1)&1\\
&16    &9.57/5&0.599(3)&0.577(1)&---&---\\
&32    &1.06/4&0.614(6)&0.572(2)&---&---\\
&48   &1.00/3&0.615(9)&0.571(3)&---&---\\
\hline
\multirow{6}{*}{$C_{R_2}$}
&4 & 6.91/6 &0.602(4) & 0.577(2)& -0.01(1) & 1 \\
&8    &6.69/5&0.605(7)&0.576(3)&-0.03(3)&1\\
&16    &1.99/4&0.63(1)&0.568(5)&-0.2(1)&1\\
&16   &7.43/5&0.599(3)&0.578(2)&---&---\\
&32   &3.40/4&0.609(6)&0.574(3)&---&---\\
&48   &0.29/3&0.62(1)&0.570(4)&---&---\\
\hline
\multirow{6}{*}{$C_{\rho L}$}
&4 & 8.61/6 & 2.22(1) & 0.576(1)& -0.94(3) & 1 \\
&8     &8.48/5&2.23(2)&0.576(2)&-0.97(9)&1\\
&16    &1.60/4&2.31(4)&0.569(3)&-1.7(3)&1\\
&16    &39.32/5&2.097(9)&0.588(1)&---&---\\
&32    &4.47/4&2.17(2)&0.580(2)&---&---\\
&48   &0.30/3&2.21(3)&0.576(3)&---&---\\
\hline
\hline
\end{tabular}
\end{center}
\end{table}

\begin{table}
\begin{center}
\caption{Fits of the two-point correlations $G_1$ and $G_2$ to Eq.~(\ref{fitg1}) and the susceptibility $\chi$ to Eq.~(\ref{fitchi1}) at the special transition.}
\label{TABLE_Gbw}
\begin{tabular}{p{0.6cm}p{0.8cm}p{0.9cm}p{1.3cm}p{1.3cm}p{1.5cm}p{0.9cm}}
\hline			
\hline
$\mathcal{Q}$ &$L_{\rm min}$ &$L_{\rm max}$ &  ${\rm chi}^2$/DoF &$a_0$&$y_h$&  $\omega_1$ \\
\hline
\multirow{10}{*}{$G_1$}
&8  &256   &2.55/4&1.54(2)&1.6888(9) &0.82(9)\\
&8 &256    &6.45/5&1.513(2)&1.6903(1)&1\\
&16   &256  &4.03/4&1.519(5)&1.6899(3)&1\\
&32  &256   &0.47/3&1.54(1)&1.6887(7)&1\\
&48  &256   &0.11/2&1.55(2)&1.688(1)&1\\
&8  &128    &1.73/3&1.53(2)&1.689(1)&0.9(1)\\
&8  &128      &3.78/4&1.512(2)&1.6904(1)&1\\
&16    &128    &2.46/3&1.517(5)&1.6900(4)&1\\
&32  &128    &0.18/2&1.53(1)&1.6889(8)&1\\
&48  &128     &0.05/1&1.54(3)&1.688(2)&1\\
\hline
\multirow{10}{*}{$G_2$}
&	8  &256   &14.01/4&1.27(2)&1.689(1)&0.9(1)\\
&8  &256   &14.60/5&1.261(2)&1.6901(2)&1\\
&16  &256   &14.20/4&1.264(5)&1.6899(4)&1\\
&32  &256   &13.50/3&1.27(1)&1.689(1)&1\\
&48  &256   &13.12/2&1.26(3)&1.690(3)&1\\
&8  &128    &4.95/3&1.27(1)&1.690(1)&0.9(1)\\
&8   &128   &5.16/4&1.261(2)&1.6901(2)&1\\
&16   &128   &4.98/3&1.263(5)&1.6900(4)&1\\
&32  &128   &4.73/2&1.27(1)&1.689(1)&1\\
&48  &128   &3.33/1&1.23(3)&1.692(3)&1\\
\hline
\multirow{10}{*}{$\chi$}
&8  &256   &3.28/4&1.45(1)&1.6894(7)&0.92(5)\\
&8  &256   &5.42/5&1.431(1)&1.6903(1)&1\\
&16  &256   &3.88/4&1.436(4)&1.6900(3)&1\\
&32  &256   &2.45/3&1.45(1)&1.6893(7)&1\\
&48 &256    &2.45/2&1.45(2)&1.689(2)&1\\
&8  &128    &0.75/3&1.44(1)&1.6895(7)&0.93(5)\\
&8   &128   &2.36/4&1.431(1)&1.6903(1)&1\\
&16   &128   &1.12/3&1.435(4)&1.6900(3)&1\\
&32   &128    &0.13/2&1.44(1)&1.6894(7)&1\\
&48  &128    &0.02/1&1.44(3)&1.690(2)&1\\
\hline
\hline
\end{tabular}
\end{center}
\end{table}

\begin{figure}
\includegraphics[height=10cm,width=8cm]{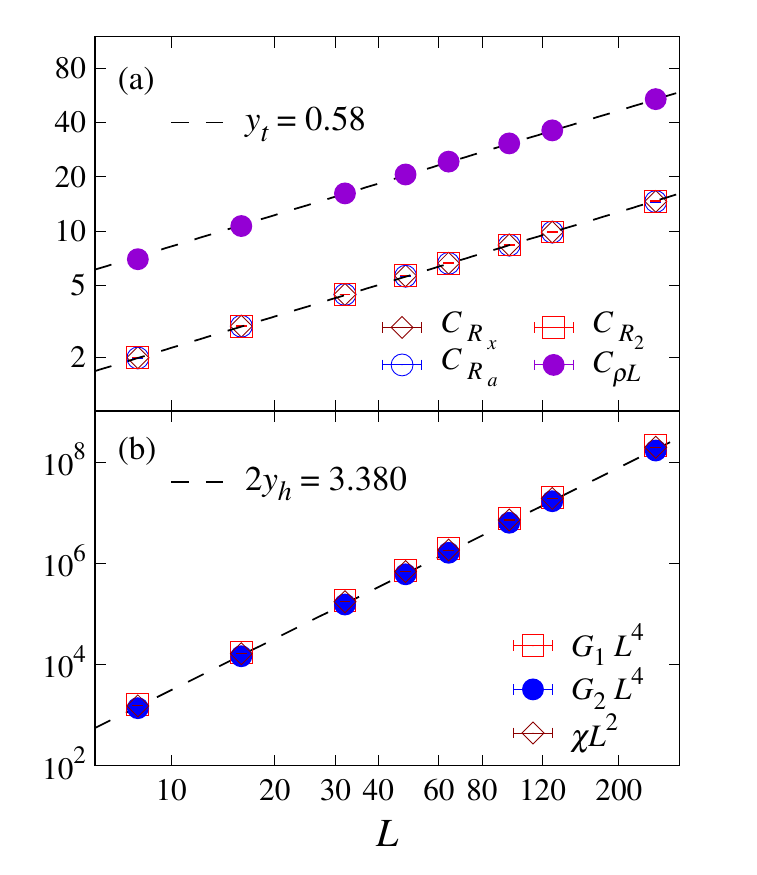}
\caption{The dependence of quantities on $L$ at the special transition. (a) Log-log plot of the covariances $C_{R_x}$, $C_{R_a}$, $C_{R_2}$ and $C_{\rho L}$ versus $L$. The slope of dashed lines stands for $y_t=0.58$. (b) Log-log plot of the scaled quantities $G_1 L^{4}$, $G_2 L^{4}$ and $\chi L^{2}$ versus $L$. The slope of dashed line stands for $2y_h=3.380$.}\label{ytyh}
\end{figure}

Assume that the SF stiffness scales as
\begin{equation}\label{3drhos}
\rho=L^{2-D} \tilde{\rho}(\epsilon L^{y_t})
\end{equation}
with $D=3$, which relates to the FSS of SF stiffness for quantum special transition~\cite{Sun2022} by $D=d+z$ ($d=2$, $z=1$). We perform fits according to the scaling ansatz
\begin{equation}\label{fitrhos1}
 \rho=L^{-1}(a_0+ a_1 \epsilon L^{y_t}+ a_2 \epsilon^2 L^{2y_t}+b_1 L^{-\omega_1}+...)
\end{equation}
with $a_0$ a constant. From Table~\ref{TABLE_S}, one finds that the estimates of $\kappa_c$ are close to those from $R_x$ and $R_a$. This finding is illustrated by Fig.~\ref{spe1}(c) demonstrating the $(\rho L)$--$\kappa$ relation, where a scale invariance point can be located at $\kappa_c \approx 0.441$, after properly handling finite-size corrections.

From the fitting results for $R_x$, $R_a$ and $\rho$, the final estimate of $\kappa_c$ is given as $\kappa_c=0.44140(9)$. Meanwhile, the estimate of $y_t$ is $y_t=0.59(4)$, which agrees with the results $0.608(4)$~\cite{deng2005surface}, $0.59(1)$~\cite{Zhang2022Surface} and $0.61(2)$~\cite{Zou2022Surface} from various contexts of O(2) special SC, yet it suffers from larger uncertainty. A more precise determination of $y_t$ will be given in the following subsection.

\subsection{Universality class}\label{fatc}
We explore the universality class for the special transition, by computing $y_t$ and $y_h$. We turn to FSS analyses right at $\kappa_c=0.44140$, which have a reduced number of fitting parameters.

To estimate $y_t$, we consider the covariances $C_O$ for dimensionless quantities ($O=R_x, R_a, R_2, \rho L$). According to Eq.~(\ref{RrGc1}),
$C_O$ scales as
\begin{equation}
C_{O}=L^{y_t} \tilde{C}_O(\epsilon L^{y_t})
\end{equation}
around $\kappa_c$. A fitting ansatz at $\kappa_c$ reads
\begin{equation}\label{fitco}
 C_{O} =L^{y_t}(a_0+b_1 L^{-\omega_1}),
\end{equation}
where $b_1 L^{-\omega_1}$ is the leading term for finite-size corrections. Log-log plots of critical covariances versus $L$ are shown in Fig.~\ref{ytyh}(a), which indicates the power-law scaling $L^{y_t}$. We perform fits to formula (\ref{fitco}), considering the situations with leading correction term ($\omega_1=1$) or without finite-size correction. The results are presented in Table~\ref{TABLE_Co}. For each of the covariances, we obtain reasonable fits in the large-size regime, even when the correction term is absent. At $L_{\rm min}=32$, we obtain $y_t=0.573(2)$, $0.572(2)$, $0.574(3)$ and $0.580(2)$ with ${\rm chi}^2/{\rm DoF} \approx 0.5$, $0.3$, $0.9$ and $1.1$, for $C_{R_x}$, $C_{R_a}$, $C_{R_2}$ and $C_{\rho L}$, respectively. Finally, from Table~\ref{TABLE_Co}, our estimate of $y_t$ is $y_t=0.58(1)$.

With $y_t=0.58$ and $\kappa_c=0.44140$, Figs.~\ref{spe1}(d), (e) and (f) display dimensionless quantities versus $(\kappa-\kappa_c)L^{y_t}$. According to Eqs.~(\ref{RrGc1}) and (\ref{3drhos}), the data collapses in the plots are indicators of reasonability for the estimated $y_t$ and $\kappa_c$.

We perform FSS analyses for the surface quantities $G_1$, $G_2$ and $\chi$, from which $y_h$ is estimated. The special transition features the power-law scaling and the critical two-point correlation obeys
\begin{equation}
g(r) \sim r^{2y_h-4}
\end{equation}
at $\kappa_c$. Hence, the FSS for $G_1$ and $G_2$ is described by
\begin{equation}\label{fitg1}
G =L^{2y_h-4}(a_0+b_1 L^{-\omega_1}).
\end{equation}
Since $\chi$ scales as $\chi=L^{2y_h-2}\tilde{\chi} (\epsilon L^{y_t})$, its FSS at $\kappa_c$ is written as
\begin{equation}\label{fitchi1}
\chi = L^{2y_h-2}(a_0+b_1 L^{-\omega_1}).
\end{equation}
The $L^{2y_h}$ divergence for scaled quantities $G_1 L^{4}$, $G_2 L^{4}$ and $\chi L^{2}$ is illustrated by Fig.~\ref{ytyh}(b). According to Eqs.~(\ref{fitg1}) and (\ref{fitchi1}), the fits for $G_1$, $G_2$ and $\chi$ are performed. The results are given in Table~\ref{TABLE_Gbw}.  The estimates for $\omega_1$ are close to $\omega_1=1$, as found in Sec.~\ref{dtc}. We note that, from each of the quantities $G_1$, $G_2$ and $\chi$, the fitting results of $y_h$ by letting $\omega_1$ be free (for smaller $L_{\rm min}$, namely $L_{\rm min}=8$) and letting $\omega_1=1$ be fixed (for larger $L_{\rm min}$, namely $L_{\rm min}=48$) are all compatible with $y_h \approx 1.690$. For $G_1$ and $\chi$, preferred fits are found with the cutoffs $L_{\rm max}=128$ and $256$. For $G_2$, precluding input data at $L=256$, which suffers from large relative statistical errors, is useful for improving the quality of fits. As a result, for $L_{\rm max}=128$, we obtain $y_h=1.690(1)$, $1.6901(2)$ and $1.6900(4)$ with ${\rm chi}^2/{\rm DoF} \approx 1.7$, $1.3$ and $1.7$, respectively. By comparing the fits in Table~\ref{TABLE_Gbw}, the final estimate of $y_h$ is $y_h=1.690(1)$.

\section{Extraordinary-log critical phase}
\subsection{Two-point correlation}\label{ss_tp}
To probe ELU, we perform extensive simulations in the deep extraordinary regime with $\kappa=5$ and $10$, and obtain precise Monte Carlo data for $G_1$ and $G_2$. According to Eq.~(\ref{tp3}), the FSS formula of $G_1$ and $G_2$ is written as
\begin{equation}\label{erLGc1}
G = a [({\rm ln}L)+c]^{-{\hat{q}}}
\end{equation}
with $c$ a non-universal constant. We perform fits for $G_1$ and $G_2$, with the results being summarized in Table~\ref{TABLE_ExtraG12}. At $\kappa=5$, the fits for $G_1$ are stable if $L_{\rm min} \gtrsim 16$, producing ${\hat{q}}=0.586(2)$ and $0.583(3)$ with ${\rm chi}^2/{\rm DoF} \approx 0.8$ and $0.6$, respectively. Comparatively, the finite-size $G_2$ data are more compatible to Eq.~(\ref{erLGc1}) for $L_{\rm min}=8$. Preferred fits with ${\rm chi}^2/{\rm DoF} \approx 1$ yield ${\hat{q}}=0.590(2)$, $0.587(3)$ and $0.582(5)$ for $L_{\rm min}=8$, $16$ and $32$, respectively. At $\kappa=10$, we obtain ${\hat{q}}=0.561(4)$, $0.566(8)$, $0.57(1)$ and $0.59(2)$ for $G_1$, as well as ${\hat{q}}=0.566(4)$, $0.564(6)$ and $0.56(1)$ for $G_2$. These estimates agree within error bars with the previous estimate ${\hat{q}}=0.59(2)$ from classical XY model~\cite{Hu2021}, providing strong evidence for the existence of ELU in OSV model.

\begin{table}
\begin{center}
\caption{Fits of the two-point correlations $G_1$ and $G_2$ to Eq.~(\ref{erLGc1}) for the extraordinary critical phase.}
\label{TABLE_ExtraG12}
\begin{tabular}{p{0.7cm}p{0.7cm}p{0.8cm}p{1.5cm}p{1.2cm}p{1.5cm}p{1.1cm}}
\hline
\hline
$\mathcal{Q}$ & $\kappa$  &$L_{\rm min}$ &${\rm chi}^2/{\rm DoF}$&$a$&$c$&${\hat{q}}$  \\
\hline
\multirow{10}{*}{$G_1$} & \multirow{5}{*}{$5$}
&8    &31.45/5&2.74(1) &5.52(2)&0.579(1)\\
&&16      &3.08/4&2.80(2) &5.65(3)&0.586(2)\\
&&32     &1.76/3&2.77(3) &5.59(6)&0.583(3)\\
&&48  &0.13/2&2.72(5) &5.50(9)&0.578(5)\\
&&64     &0.11/1&2.72(7) &5.5(1)&0.577(8)\\
&\multirow{5}{*}{$10$}
&8     &43.83/5&3.60(4) &10.60(8)&0.541(3)\\
&&16      &2.61/4&3.87(6) &11.1(1)&0.561(4)\\
&&32     &1.93/3&4.0(1) &11.3(2)&0.566(8)\\
&&48     &1.54/2&4.0(2) &11.5(4)&0.57(1)\\
&&64     &0.46/1&4.3(3) &11.9(6)&0.59(2)\\
\hline
\multirow{10}{*}{$G_2$}  & \multirow{5}{*}{$5$}
&8    &6.22/5&2.84(2) &6.09(3)&0.590(2)\\
&&16    &3.15/4&2.81(2) &6.02(5)&0.587(3)\\
&&32     &1.38/3&2.76(4) &5.93(8)&0.582(5)\\
&&48     &0.24/2&2.71(7) &5.8(1)&0.576(7)\\
&&64     &0.15/1&2.7(1) &5.8(2)&0.57(1)\\
&\multirow{5}{*}{$10$}
&8  &5.46/5&3.95(6) &11.6(1)&0.566(4)\\
&&16   &5.27/4&3.92(8) &11.6(2)&0.564(6)\\
&&32   &5.16/3&3.9(1) &11.5(3)&0.56(1)\\
&&48   &4.13/2&4.1(2) &11.9(4)&0.57(2)\\
&&64  &0.12/1&4.6(5) &12.9(7)&0.61(3)\\
\hline
\hline
\end{tabular}
\end{center}
\end{table}

\begin{table}
\begin{center}
\caption{Fits of the susceptibility $\chi$ to Eq.~(\ref{erLGc2}) for the extraordinary critical phase.}
\label{TABLE_ExtraChi}
\begin{tabular}{p{0.7cm}p{1.0cm}p{1.55cm}p{1.5cm}p{1.5cm}p{1.1cm}}
\hline	
\hline		
$\kappa$ & $L_{\rm min}$  &${\rm chi}^2/{\rm DoF}$ & $a$ & $c$ & ${\hat{q}}$ \\
\hline
 \multirow{4}{*}{$5$}
&16    &23.43/4&2.93(2) &6.06(3)&0.600(2)\\
&32    &3.35/3&2.80(3) &5.82(6)&0.586(4)\\
&48     &1.45/2&2.75(5) &5.7(1)&0.580(5)\\
&64     &1.39/1&2.74(8) &5.7(1)&0.579(8)\\
\hline
 \multirow{4}{*}{$10$}
&16  &5.16/4&4.17(7) &11.9(1)&0.580(5)\\
&32   &4.35/3&4.1(1) &11.7(2)&0.574(8)\\
&48   &2.68/2&4.3(2) &12.1(4)&0.59(1)\\
&64   &0.002/1&4.7(4) &12.7(6)&0.61(2)\\
\hline
\hline
\end{tabular}
\end{center}
\end{table}

From Eq.~(\ref{tp3}), we obtain a FSS formula for $\chi$, which reads
\begin{equation}\label{erLGc2}
\chi =a L^2 [({\rm ln}L)+c]^{-\hat{q}},
\end{equation}
due to $\chi \sim \int g(r) r {\rm d}r$. Table~\ref{TABLE_ExtraChi} displays the existence of preferred fits to Eq.~(\ref{erLGc2}) for $\kappa=5$ and $10$. For $\kappa=5$, we have the fitting results ${\hat{q}}= 0.586(4)$, $0.580(5)$ and $0.579(8)$ with ${\rm chi}^2/{\rm DoF} \approx 1.1$, $0.7$ and $1.4$, for $L_{\rm min}=32$, $48$ and $64$, respectively. For $\kappa=10$, we obtain ${\hat{q}}= 0.580(5)$, $0.574(8)$ and $0.59(1)$ with ${\rm chi}^2/{\rm DoF} \approx 1.3$, $1.5$ and $1.3$, for $L_{\rm min}=16$, $32$ and $48$, respectively. Therefore, the estimates of ${\hat{q}}$ from $\chi$ are compatible with the results from $G_1$ and $G_2$.

\begin{figure}
\includegraphics[height=10cm,width=8cm]{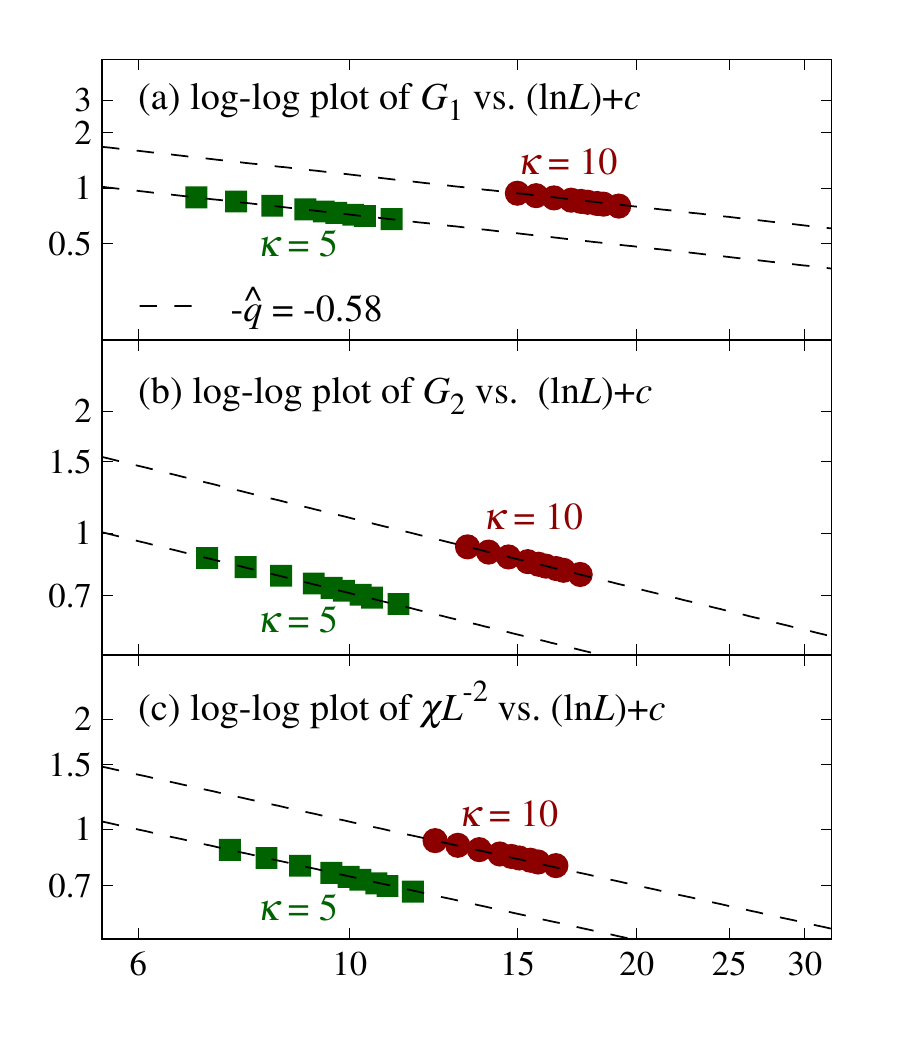}
\caption{Log-log plot of $G_1$ (a), $G_2$ (b) and $\chi L^{-2}$ (c) versus $({\rm ln}L)+c$. The dashed lines stand for the critical scaling $({\rm ln}L)^{-\hat{q}}$ with $\hat{q}=0.58$. The constant $c$ is non-universal and comes from the preferred least-squares fits of $G_1$ and $G_2$ to Eq.~(\ref{erLGc1}) or $\chi$ to Eq.~(\ref{erLGc2}).}\label{hatq}
\end{figure}

Generally speaking, the FSS analysis involving ${\rm ln}L$ is difficult. Hence, the stability of fits is examined by varying $L_{\rm min}$ and we do not trust any single fit even though the chi-squared criterion is satisfied. The estimates of fitting parameters (including $\hat{q}$) arise from a comparison of the fits with different $L_{\rm min}$. Moreover, to monitor the corrections to scaling, we systematically compare the estimates of $\hat{q}$ from various quantities. We also compare the results from different interaction strengths in the extraordinary-log regime. A similar procedure was applied in a previous study~\cite{Hu2021}, of which the estimate of $\hat{q}$ has been confirmed by independent studies in various contexts (Table~\ref{tb1}). Here, by comparing the preferred fits for $G_1$, $G_2$ and $\chi$, we estimate $\hat{q}=0.58(2)$. By adopting the parameter $c$ from the fits, we plot $G_1$, $G_2$ and $\chi L^{-2}$ versus $({\rm ln}L)+c$ in Fig.~\ref{hatq}, which illustrates mutually consistent results for universal and non-universal parameters.

\begin{figure}
\includegraphics[height=6cm,width=8cm]{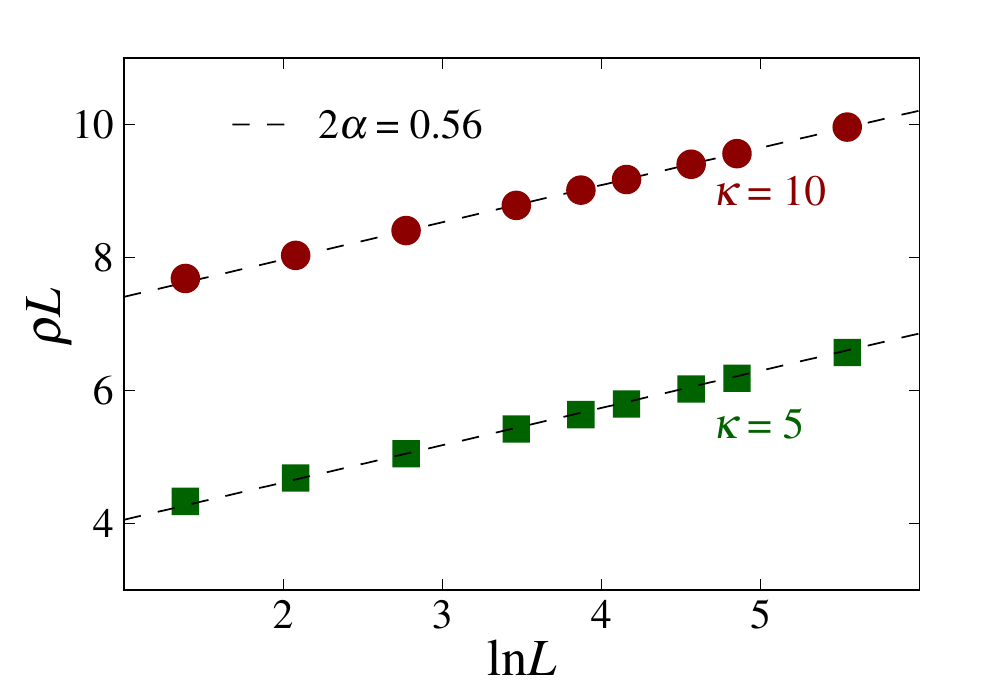}
\caption{The scaled SF stiffness $\rho L$ versus ${\rm ln}L$. The dashed lines stand for $2\alpha$.}\label{alpha}
\end{figure}

\begin{table}
\begin{center}
\caption{Fits of the SF stiffness $\rho$ to Eq.~(\ref{rLGc3}) for the extraordinary critical phase. The underlines denote that the data for $L=128$ are not included in fitting.}
\label{TABLE_Extrarho}
\begin{tabular}{p{0.8cm}p{1.05cm}p{1.05cm}p{1.75cm}p{1.7cm}p{1.15cm}}
\hline		
\hline
$\kappa$&$L_{\rm min}$ &$L_{\rm max}$ &${\rm chi}^2/{\rm DoF}$ &$\alpha$  &$b$ \\
\hline
 \multirow{10}{*}{$5$}
&32  &256    &32.11/4&0.2770(4) &3.498(3) \\
&48   &256   &5.38/3&0.2785(5) &3.484(4) \\
&64   &256  &3.87/2&0.2790(6) &3.479(6) \\
&96  &256  &0.79/1&0.280(1)& 3.47(1) \\
&32  &128 &18.79/3 &0.2756(5) &3.509(4)\\
&48  &128 &3.15/2 &0.2776(7) &3.491(6) \\
&64  &128 &2.91/1&0.278(1) &3.49(1) \\
&\underline{32}  &\underline{256}    &32.08/3&0.2769(4) &3.499(4) \\
&\underline{48}   &\underline{256}   &5.38/2&0.2785(5) &3.484(5) \\
&\underline{64}   &\underline{256}  &3.87/1&0.2790(7) &3.479(6) \\
\hline
 \multirow{10}{*}{$10$}
&32  &256  &7.35/4&0.2822(6) &6.827(5) \\
&48  &256  &6.28/3&0.2828(8) &6.821(8) \\
&64  &256  &4.13/2&0.284(1) &6.81(1) \\
&96  &256  &3.51/1&0.285(2) &6.80(1) \\
&32   &128  &1.13/3&0.2810(8) &6.837(7) \\
&48   &128  &1.10/2&0.281(1) &6.84(1) \\
&64   &128  &1.03/1&0.281(2) &6.83(2) \\
&\underline{32}  &\underline{256}    &3.23/3&0.2828(7) &6.823(6)\\
&\underline{48}   &\underline{256}   &1.85/2&0.2835(9) &6.816(8) \\
&\underline{64}   &\underline{256}  &0.05/1&0.284(1) &6.81(1) \\
\hline
\hline
\end{tabular}
\end{center}
\end{table}

\subsection{Superfluid stiffness}
We examine the analogy of $\rho$ to the SF stiffness of open-edge Bose-Hubbard model considered in Ref.~\cite{Sun2022}, where it was defined through the winding number fluctuations in path-integral world-line representation. As shown in Fig.~\ref{alpha}, there is a linear divergence of $\rho L$ on ${\rm ln}L$. The renormalization-group universal parameter $\alpha$ controls the FSS of $\rho$, which can be written as
\begin{equation}~\label{rLGc3}
\rho L = 2\alpha ({\rm ln}L)+b.
\end{equation}
Estimates of $\alpha$ come from the fits of $\rho$ to Eq.~(\ref{rLGc3}), which are summarized in Table~\ref{TABLE_Extrarho}. For $\kappa=5$,
we obtain $\alpha=0.2785(5)$, $0.2790(6)$ and $0.280(1)$ with ${\rm chi}^2/{\rm DoF} \approx 1.8$, $1.9$ and $0.8$, for $L_{\rm min}=48$, $64$ and $96$, respectively. We are aware of the price of including large-size data with relatively large uncertainties, and also perform fits with $L=256$ being precluded, i.e., $L_{\rm max}=128$. As a result, we obtain $\alpha=0.2776(7)$ and $0.278(1)$ with ${\rm chi}^2/{\rm DoF} \approx 1.6$ and $2.9$, respectively. Then, we perform fits with the second-largest size $L=128$ being precluded yet $L=256$ being contained, for which the residuals are larger. A similar fitting procedure is applied to $\kappa=10$, and preferred fits are achieved. For $L_{\rm max}=128$, we obtain $0.2810(8)$, $0.281(1)$ and $0.281(2)$ with ${\rm chi}^2/{\rm DoF} \approx 0.4$, $0.6$ and $1.0$, respectively. When $L=128$ is precluded yet $L=256$ is contained, we obtain $0.2828(7)$ and $0.2835(9)$ with ${\rm chi}^2/{\rm DoF} \approx 1.1$ and $0.9$, respectively. Therefore, the estimates of $\alpha$ from $\kappa=5$ and $10$ are close to each other. By comparing the fitting results in Table~\ref{TABLE_Extrarho}, the universal value of $\alpha$ is estimated to be $\alpha=0.28(1)$.

\begin{figure}
\includegraphics[height=8cm,width=8cm]{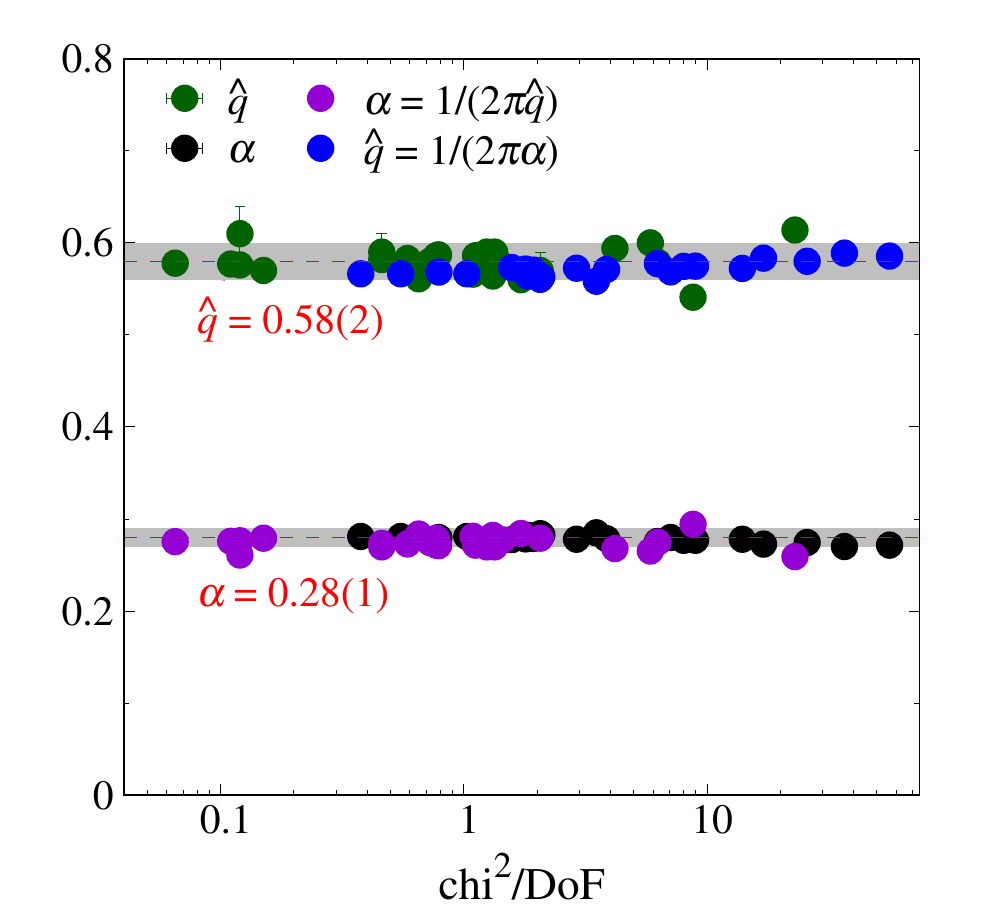}
\caption{Verification for the scaling relation~(\ref{qp2}) with $N=2$. The green and black symbols with error bars stand for fitting results from FSS analyses, while the purple and blue symbols without error bars are transformed from the fitting results via the relation $\alpha \hat{q}=1/(2\pi)$. The shadowed areas centered at dashed red lines denote the critical exponent $\hat{q}=0.58(2)$ and the renormalization-group parameter $\alpha=0.28(1)$.}\label{scaling}
\end{figure}

\subsection{Scaling relation}
We proceed to verify the scaling relation~(\ref{qp2}) of the critical exponent $\hat{q}$ and the renormalization-group parameter $\alpha$. Figure~\ref{scaling} demonstrates the fitting results for $\hat{q}$ and $\alpha$ versus ${\rm chi}^2/{\rm DoF}$, which are quoted from Tables~\ref{TABLE_ExtraG12}, \ref{TABLE_ExtraChi} and \ref{TABLE_Extrarho}. In the plot, the two shadowed areas with $\hat{q}=0.58(2)$ and $\alpha=0.28(1)$ denote the final estimates from fitting. Next, using scaling relation~(\ref{qp2}) with $N=2$, namely $\alpha \hat{q}=1/(2\pi)$, we obtain estimates of $\alpha$ and $\hat{q}$ from each other, and the results are also presented in Fig.~\ref{scaling}. It is found that the estimates of $\hat{q}$ and $\alpha$ from scaling relation are close to the final estimates indicated by shadowed areas, particularly when ${\rm chi}^2/{\rm DoF} \approx 1$ is approached. Hence, the scaling relation~(\ref{qp2}) is compatible with present numerical results.

\section{Discussion}
To bridge the recent observations of exotic SC in classical statistical mechanical models~\cite{toldin2020boundary,Hu2021,ToldinMetlitski2021extraordinary} and quantum Bose-Hubbard model~\cite{Sun2022}, we formulate the OSV model for special and extraordinary-log criticality, which is extensively simulated by a worm Monte Carlo algorithm. For the special transition, the thermal and magnetic renormalization exponents are estimated to be $y_t=0.58(1)$ and $y_h=1.690(1)$ respectively, which are consistent with recent results from classical spin models of emergent O(2) criticality~\cite{Zhang2022Surface,Zou2022Surface}. For the extraordinary-log phase, the critical exponent $\hat{q}$ and the universal renormalization-group parameter $\alpha$ are estimated to be $\hat{q}=0.58(2)$ and $\alpha=0.28(1)$, which are compatible with scaling relation (\ref{qp2}) with $N=2$. Meanwhile, the estimated $\hat{q}$ and $\alpha$ are fully consistent with previous results from XY model~\cite{Hu2021}. Moreover, the SF stiffness scales as $L^{-1}$ at the special transition and as $L^{-1} ({\rm ln}L)$ for the extraordinary-log critical phase. These features resemble the scaling formulae of SF stiffness for open-edge quantum Bose-Hubbard model~\cite{Sun2022}, where the stiffness was sampled over world-line configurations. Hence, the present work provides an alternative demonstration for ELU and bridges recent numerical observations over classical and quantum SC. As a byproduct, it is promising that the quantitative results for special and extraordinary-log criticality would serve as a long-standing benchmark.

One direction for future work may be to finely tune the geometries of boundaries
for a critical Bose-Hubbard or Villain system by employing a full Suzuki-Trotter-type limiting procedure that underlies the quantum-classical correspondence. Such an activity would offer a routine to reconcile the current questions about SC in dimerized quantum antiferromagnets~\cite{zhang2017unconventional,ding2018engineering,Weber2018,weber2019nonordinary,jian2021continuous,Weber2021,Zhu2021Exotic,ding2021special}, where the emergence of SC subtly depends on geometric settings of boundaries and relates to symmetry-protected topological phases.

\begin{acknowledgments}
One of us (J.P.L.) wishes to thank Youjin Deng and Minghui Hu for the collaboration in an earlier study~\cite{Hu2021}. The present work has been supported by the National Natural Science Foundation of China (under Grant Nos. 12275002, 11975024, and
11774002) and the Education Department of Anhui.
\end{acknowledgments}

\bibliography{papers}

\end{document}